\documentclass[12pt, notitlepage, usenatbib]{article}
\usepackage[margin = .75in]{geometry}
\usepackage[doublespacing]{setspace}
\usepackage{graphicx}
\usepackage{titling}
\usepackage[pagewise]{lineno}
\usepackage{amsmath, bm, verbatim}
\usepackage{hyperref}
\usepackage{xcolor}
\usepackage{listings}
\usepackage{cite}

\definecolor{darkgreen}{rgb}{0.0, 0.5, 0.0}  
\definecolor{blue}{rgb}{0.0, 0.0, 1.0}       

\usepackage{comment}

\usepackage[singlelinecheck=false ]{caption}

\title{Combining missing data imputation and internal validation in clinical risk prediction models}
\preauthor{}
\postauthor{}
\author{}
\predate{}
\postdate{}
\date{}

\usepackage{authblk}

\author[1]{Junhui Mi}
\author[2]{Rahul D. Tendulkar}
\author[3]{Sarah M. C. Sittenfeld}
\author[1]{Sujata Patil}
\author[1]{Emily C. Zabor}

\affil[1]{Department of Quantitative Health Sciences, Cleveland Clinic Research, Ohio, USA}
\affil[2]{Department of Radiation Oncology, Taussig Cancer Institute, Cleveland Clinic, Ohio, USA}
\affil[3]{Department of Radiation Oncology, The Barrett Cancer Center, University of Cincinnati, Ohio, USA}

\begin{document}


\begin{titlingpage}
\maketitle

\begin{abstract}
\normalsize
Methods to handle missing data have been extensively explored in the context of estimation and descriptive studies, with multiple imputation being the most widely used method in clinical research. However, in the context of clinical risk prediction models, where the goal is often to achieve high prediction accuracy and to make predictions for future patients, there are different considerations regarding the handling of missing data. As a result, deterministic imputation is better suited to the setting of clinical risk prediction models, since the outcome is not included in the imputation model and the imputation method can be easily applied to future patients. In this paper, we provide a tutorial demonstrating how to conduct bootstrapping followed by deterministic imputation of missing data to construct and internally validate the performance of a clinical risk prediction model in the presence of missing data. Extensive simulation study results are provided to help guide decision-making in real-world applications.

\bigskip
\noindent
\textbf{Keywords}: prediction model; imputation; deterministic imputation; multiple imputation; missing data; risk prediction
\end{abstract}
\end{titlingpage}


\setstretch{2.1}

\section{Introduction}\label{sec1}

When building clinical risk prediction models, it is common to encounter missing covariate data. Methods to handle missing data have been extensively explored in the context of estimation and descriptive studies \cite{Rubin1976, Rubin1996, Marshall2009, Marshall2010, Austin2021, Buuren2011, Little2012}, but there are different considerations regarding missing data in the context of prediction \cite{Sperrin2020}. With description and estimation, the main focus of missing data methods has been on overcoming bias in parameter estimates, and ensuring variability is properly addressed so that inference will be as accurate as possible. But in the context of prediction, interest is not necessarily on estimation or precision at all, but rather on prediction accuracy and creating prediction models that can be usefully applied to data on future patients. In addition, conditions under which a clinical prediction model will be deployed, especially the unknown nature of the outcome at the time of prediction, must be considered when determining the best approach to handle missing data. 

Multiple imputation has been the gold standard to handle missing data in clinical research for some time now \cite{Rubin1996}. Stochastic imputation methods such as multiple imputation are focused on adjusting the variability of estimates to account for the uncertainty in the imputation modeling process. Stochastic imputation requires incorporating the outcome in the imputation model to ensure unbiased results \cite{DAgostinoMcGowan2024}. For these reasons, stochastic imputation is not well suited to the context of clinical prediction models. Deterministic imputation, also known as regression imputation, fits a single imputation model and directly replaces missing values with predicted values in a fixed, error-free manner. The outcome should not be included in the setting of deterministic imputation, as including the outcome will introduce bias in the estimated regression coefficients \cite{DAgostinoMcGowan2024}. In addition, deterministic imputation approaches can be applied to future patients using both the imputation models and the full prediction model. In contrast, multiple imputation approaches typically would require access to the full development dataset in order to be applied to a future patient, and even then could not be used in the context of unknown outcome values \cite{FletcherMercaldo2020}. Deterministic imputation has been shown to perform at least as well as multiple imputation in deploying clinical prediction models, provided the outcome is appropriately excluded from the imputation model. \cite{Sisk2023}. 

Another consideration in the context of clinical prediction model development is the need for internal validation. While external model validation has often been considered the gold standard for performance assessment of clinical risk prediction models, it is common in practice to only have access to a single dataset during the development process. In this context, internal validation that encompasses the entire model-building process, including variable selection, is recommended. When internal validation is done with bootstrapping, it can even outperform external validation in some settings \cite{Steyerberg2016}. The next logical question is the order in which to perform bootstrapping and imputation. In the setting where estimation is of interest, it has been shown that bootstrapping prior to multiple imputation achieves superior confidence interval coverage \cite{Bartlett2020, Schomaker2018}. In the context where prediction is of interest, it could be argued that bootstrapping prior to imputation is the only acceptable approach, since doing imputation prior to bootstrapping could be considered using information from the development process in the validation process. 

In this paper, we provide a tutorial demonstrating how to conduct bootstrapping followed by deterministic regression imputation of missing data to construct and internally validate the performance of a clinical risk prediction model. Code examples in the R programming language are included throughout \cite{RCT2024}. We also provide simulation results to guide decision-making when considering building a clinical risk prediction model in the presence of missing predictor data.

\section{Performance metrics}\label{sec2}

We evaluated traditional performance measures used in clinical prediction models, incorporating resampling methods to correct for overfitting and obtain more robust performance estimates. Specifically, we assessed the model’s discriminative ability using the area under the receiver operating characteristic curve (AUC) and measured overall model performance with the Brier score \cite{Steyerberg2010}. We reported both apparent (uncorrected) AUC and Brier scores and bias-corrected estimators, as detailed below.

The AUC represents the probability that a randomly chosen positive instance (event) ranks higher in predicted risk than a randomly chosen negative instance (non-event), with an AUC of 1 indicating perfect discrimination \cite{Wahl2016}. For time-to-event outcomes, the AUC is adapted to a time-dependent AUC, which evaluates the model’s discriminative capacity at specific time points. The Brier score assesses the accuracy of risk predictions by calculating the mean squared difference between predicted probabilities and observed outcomes, where a Brier score of 0 indicates perfect accuracy \cite{Wahl2016}. For survival analysis, the time-dependent Brier score is used, calculating the difference between predicted survival probabilities and actual outcomes at specific time points.

\subsection{Apparent}

The apparent estimator measures predictive performance directly on the original sample, often yielding overly optimistic results as the model is evaluated on the same data used for its development. This can lead to underestimation of the true prediction error \cite{Iba2021}.

\subsection{Bootstrap-corrected estimator}

The bootstrap-corrected estimator adjusts the apparent estimator by accounting for optimism, defined as the average difference between the performance on bootstrap samples and the original dataset \cite{Harrell1996}. By subtracting this optimism, the corrected estimator provides a more realistic assessment of model performance, accounting for overfit. However, because approximately 63.2\% of the original sample is typically included in each bootstrap sample, this overlap can still lead to a slight overestimation of performance \cite{Iba2021}.

\subsection{.632 estimator}

The .632 bootstrap method addresses the potential optimistic bias of conventional bootstrap validation by weighting performance on in-sample data at 0.632 and out-of-sample, or test, data at 0.368 \cite{Efron1983}. This weighting can produce a more balanced estimate.

\subsection{.632+ estimator}

As an advancement of the .632 method, the .632+ estimator further adjusts for overfit models by dynamically changing the weights based on an estimate of the amount of overfit in the model \cite{Efron1997}. This approach is particularly suitable for complex models, as it optimally balances in-sample and out-of-sample performance, making it ideal for high-risk overfitting scenarios. When overfitting is minimal, the .632+ and .632 estimators yield similar results. Calculating the .632+ estimator requires selecting a benchmark value representing a non-informative model. For the AUC, this value is 0.5. For the Brier score, a value of 0.25 represents a non-informative model where one predicts a 50\% risk for all participants \cite{Steyerberg2010, Gerds2021}.

\section{Data generation}\label{sec3}

\subsection{Motivating dataset}

This study is motivated by a case study where interest was in the association between post-mastectomy radiation therapy (PMRT) and breast cancer outcomes, adjusted for a variety of other patient and disease characteristics \cite{Sittenfeld2022}. In the original study, there was interest in both estimation and prediction. Eleven covariates were selected \emph{a priori} for inclusion in the prediction model. The primary predictor of interest, PMRT, had no missing data. Six of the other covariates had missing data, with four missing in $<3\%$ of patients and two missing in $>15\%$ of patients. Seventy-eight percent of patients had complete covariate data, $21\%$ were missing one covariate, $1.2\%$ were missing two covariates, and $0.03\%$ (n=1) were missing three covariates. The original study employed multiple imputation of missing data values followed by bootstrapping for interval validation. The dataset consisted of 3,532 patients, of whom 2,750 had complete data. Median follow-up time among survivors was 6.8 years (Range: 0.4-13.8). During that time, 626 patients died from any cause.

\subsection{Simulated data generation}

Simulated data will be generated for use throughout this paper based on the observed distributions of covariates and associations between covariates and outcome in the previously described motivating dataset \cite{Sittenfeld2022}. While the original study investigated a variety of clinical outcomes, some in the context of competing risks, here focus is only on associations with overall survival. Overall survival is a censored time-to-event outcome, with time measured from the date of diagnosis to the date of death. Patients who have not yet died are censored at the last date they were known to be alive. 

Death and censoring times were generated using the Weibull distribution. The baseline shape and scale parameters for the distribution of death times were estimated using a parametric Weibull model fit to the complete case dataset including all 11 covariates, resulting in a shape parameter of $1.6$ and a scale parameter of $122$. The baseline shape and scale parameters for the distribution of censoring times were estimated using a parametric Weibull model fit to the full dataset with no covariates, resulting in a shape parameter of $2.6$ and a scale parameter of $8.2$. True death times were generated from a Weibull distribution using the inverse uniform transform method. First, a uniform random variable, $u$, was generated. Then, Weibull death times were calculated as $t_i = b \times ((-log(u_i))/ \exp(\boldsymbol{\beta'} \boldsymbol{X_i}))^{(1/a)}$, where $a$ is the Weibull shape parameter, $b$ is the Weibull scale, $\boldsymbol{\beta'}$ is the vector of coefficients for the covariates, $\boldsymbol{X_i}$ is the vector of covariate values, $(X_{i1}, \ldots, X_{ip})$, for covariate $j = 1 \ldots p$ and participant $i = 1 \ldots n$. Censoring times $c_i$ were simulated directly from a Weibull distribution. Observed overall survival time for each participant is $s_i = min(t_i, c_i)$. The event, $\delta_i$, is death (1) if $t_i \leq c_i$ and censored (0) if $t_i > c_i$.

The true hazard ratios and type for each covariate are shown in Table~\ref{tab1}. Covariate values were generated from a multivariate normal distribution, using mean vector $(0.6145, 57.6495, \allowbreak  2.3665, 0.4225, 0.1538, 0.2421, 0.4142, 0.8680, 0.1695, 0.1636, 0.8891)$ for $(X_1, \ldots, X_{11})$, respectively, and the covariance matrix shown in Table~\ref{tab2}, both estimated from the original dataset. Binary variables were dichotomized at 0.5. 

\subsection{Missing data generation}

Missingness was imposed using a missing at random framework, where missingness of a given covariate may depend  on observed values but occurs independently of any unobserved values, following the approach of Marshall et al\cite{Marshall2010} and Wahl et al\cite{Wahl2016}. Let $X_{ij}$ denote the $j$th covariate for observation $i$, with $i = 1, \ldots , n$ and $j = 1, \ldots , p$. $M_{ij}$ denotes the indicator for its missingness, $M_{ij} = I(X_{ij} \text{ missing})$. Then, the probability of missingness for each covariate value was modeled as a function of the value of one other covariate and the missingness status of one other covariate as
$$P(M_{ij}=1) = logistic(\gamma_{0j} + \gamma_{1j} M_{ik_j} + \gamma_{2j} X_{il_j})$$ 
\noindent
where $M_{ik_j}$ denotes the missingness of a randomly chosen covariate and $X_{il_j}$ denotes the observed value of a randomly chosen other covariate, such that $k_j \neq l_j$.

$\gamma_{1j}$ was defined as
\[
    \gamma_{1j} = 
\begin{cases}
    0,& \text{if } j=1\\
    \log(OR_{jk_j}),& \text{otherwise}
\end{cases}
\]
\noindent
$OR_{jk_j}$ was calculated based on a cross-tabulation of the missing proportions for $X_{ij}$ and $X_{ik_j}$. Let $p_{rc}$ denote the cell proportion of missingness for the cross tabulation of $M_{ij}$ in the row so that $r$ takes the value 0 if $M_{ij}$ is not missing and the value of 1 if $X_{ij}$ is missing, and $X_{ik_j}$ in the column so that $c$ takes the value of 0 if $X_{ik_j}$ is not missing and the value of 1 if $X_{ik_j}$ is missing. Then $OR_{jk_j} = \frac{p_{00} \times p_{11}}{p_{10} \times p_{01}}$. 

The intercepts $\gamma_{0j}$ were estimated by solving the equation 
$$\gamma_{0j} = \log \left( \frac{P(M_{ij}=1)}{1-P(M_{ij}=1)} \right) - \gamma_{1j} P(M_{ik_j}=1) - \gamma_{2j} \bar{X}_{il_j}$$.

The specific missing data mechanisms for each covariate with potential missingness, and the fixed values of $\gamma_{2j}$, are shown in Table~\ref{tab3} .

\subsection{Missing data imputation}\label{sec3d}

The deterministic regression imputation method uses a set of complete predictor variables to generate a single predicted value for each missing entry in the target column.\cite{Buuren2011, Buuren2018} In deterministic regression imputation involving multiple missing variables, there are two key strategies: using separate imputation models for each variable or employing a sequential imputation approach. Sequential imputation accounts for potential dependencies between the missing variables, but inaccuracies in imputing one variable can introduce errors in subsequent imputations. This method is commonly used in stochastic or multiple imputation frameworks, which typically repeat the process across several cycles.\cite{White2011} In contrast, separate imputation models are better suited when relationships between missing variables are unclear, as they prevent the propagation of uncertainty from one poorly predicted variable to the next. In our analysis, we employed deterministic regression imputation using independent models for each variable to mitigate potential errors.

For each missing variable, a generalized linear model was fit with the missing variable as the outcome and all other complete variables as the predictors, in the subset of patients not missing the outcome variable. The link function was logit for binary missing variables and identity for continuous missing variables. Next, predicted response values for the missing variable were generated to serve as imputed values. Three imputation strategies were explored: (1) imputing all missing values, (2) imputing only for variables missing across $>10\%$ of participants, and (3) imputing only for participants missing two or fewer variables.

\section{Guided example}\label{sec4}

The guided example is meant to mirror the setting where we have a single clinical dataset that contains missing values for some predictors, and the goal is to construct and internally validate a clinical risk prediction model. In this setting, the ``truth'', based on full data with no missingness, is unknown. The guided example considers the setting with missingness in three covariates $X_1$, $X_3$, and $X_4$ such that $P(M_{i1}=1) = 0.05$, $P(M_{i3}=1) = 0.15$, and $P(M_{i4}=1) = 0.30$. The joint missingness proportions were $P(M_{i1}=1 \text{ \& } M_{i3}=1) = 0.02$ and $P(M_{i3}=1 \text{ \& } M_{i4}=1) = 0.075$. The other eight covariates are complete. In line with the motivating example, the outcome of interest is a time-to-event outcome, and the data are analyzed using multivariable Cox regression. The R code to generate and analyze the single synthetic dataset used in this section, as well as the data file itself, is available on GitHub at \url{https://github.com/zabore/manuscript-code-repository/tree/master/Mi-Zabor_bootstrap-impute-predict-tutorial}. Before running the code in this section, either load the data file or run the code to generate the dataset in order to have access to the data frame object \verb|dat0| used in the guided example. Here, the prediction horizon is set to 5 years, and 500 bootstrap samples were generated. 

First, load the R packages needed for this analysis \cite{RCT2024, Wickham2023, Wickham2023a, Gerds2023, Gerds2021, survival-package, survival-book}.

\begin{lstlisting}[language=R]
library(purrr)
library(purrr)
library(dplyr)
library(riskRegression)
library(survival)
\end{lstlisting}

\noindent
To perform imputation on both the original dataset and each bootstrap sample separately, we incorporate the imputation code within a function. The first step fits the imputation models with each variable with missing data as the outcome, in the subset of patients not missing that variable. The second step generates a predicted value for each variable with missing data for all patients in the dataset, in line with the first imputation method described in Section~\ref{sec3d}. The final step keeps the predicted value for patients missing the variable of interest and keeps the original value for patients not missing the variable of interest.

\begin{lstlisting}[language=R]
do_imp <- function(dataset) {
  # Imputation models
  imp_x1 <- glm(x1 ~ x2 + x5 + x6 + x7 + x8 + x9 + x10 + x11,
    data = dataset, family = "binomial")
  imp_x3 <- glm( x3 ~ x2 + x5 + x6 + x7 + x8 + x9 + x10 + x11,
    data = dataset, family = "gaussian")
  imp_x4 <- glm(x4 ~ x2 + x5 + x6 + x7 + x8 + x9 + x10 + x11,
    data = dataset,  family = "binomial")
  # Predicted value
  x1_pred <- ifelse(predict(imp_x1, newdata = dataset, 
    type = "response") > 0.5, 1, 0)
  x3_pred <- predict(imp_x3, newdata = dataset)
  x4_pred <- ifelse(predict(imp_x4, newdata = dataset, 
    type = "response") > 0.5, 1, 0)
  # Combine observed and predicted
  dat <- dataset |> mutate(
    x1_imp = ifelse(is.na(x1), x1_pred, x1),
    x3_imp = ifelse(is.na(x3), x3_pred, x3), 
    x4_imp = ifelse(is.na(x4), x4_pred, x4))
  return(dat)
}
\end{lstlisting}

\noindent
Use the \verb!do_imp()! function to impute directly into the original dataset \verb|dat0|, creating a new data frame object \verb|dat|:

\begin{lstlisting}[language=R]
# Impute into the original dataset
dat <- do_imp(dat0)
\end{lstlisting}

\noindent
Next, generate 500 bootstrap samples with replacement, and impute into each bootstrap sample:
\begin{lstlisting}[language=R]
# set seed for replication
set.seed(20240819)
# Take 500 bootstrap samples of the data
boot_dat <- map(1:500, ~ slice_sample(dat), prop = 1, replace = TRUE))
# Then impute into the bootstrap samples
boot_imp_dat <- boot_dat |>  map(~ do_imp(.x))
\end{lstlisting}

\noindent
Determine the apparent performance by fitting a multivariable Cox model to the original imputed dataset and evaluating its performance through AUC and Brier score:
\begin{lstlisting}[language=R]
# Fit a model to the original imputed data
app_mod <- coxph(Surv(s, delta) ~ x1_imp + x2 + x3_imp + x4_imp + 
    x5 + x6 + x7 + x8 + x9 + x10 + x11, data = dat,
    x = TRUE, y = TRUE)
# Calculate performance in original imputed data
app_auc_brier <- Score(list("fit" = app_mod), 
    formula = app_mod[["formula"]], data = dat, 
    se.fit = FALSE, conf.int = FALSE, times = 5)
# Extract apparent AUC and Brier score
app_auc <- app_auc_brier[["AUC"]][["score"]][["AUC"]]
app_brier <- app_auc_brier[["Brier"]][["score"]][["Brier"]][[2]]
\end{lstlisting}

\noindent
The apparent AUC is \verb|app_auc=0.722| and the apparent Brier score is \verb|app_brier=0.092|. Then, calculate the bootstrap-corrected performance by fitting a multivariable Cox regression model to each bootstrap sample, then evaluate its performance in both the bootstrap sample and the original imputed dataset. The bootstrap-corrected performance is calculated as the apparent performance minus the average difference between the performance in each bootstrap sample and the performance in the original imputed data:
\begin{lstlisting}[language=R]
# Fit a model to each bootstrap sample
b_mod <- map(boot_imp_dat, ~ coxph(app_mod[["formula"]], 
    data = .x, x = TRUE, y = TRUE)) 
# Calculate performance in each bootstrap sample
b_auc_brier <- map2(boot_imp_dat, b_mod, ~ Score(list("b_fit" = .y), 
    formula = app_mod[["formula"]], data = .x, se.fit = FALSE,
    conf.int = FALSE, times = 5))
b_auc <- map_dbl(b_auc_brier, ~ .x[["AUC"]][["score"]][["AUC"]])
b_brier <- map_dbl(b_auc_brier, ~ .x[["Brier"]][["score"]][["Brier"]][[2]])
# Calculate performance in original data
o_auc_brier <- map(b_mod, ~ Score(list("ofit" = .x), 
    formula = app_mod[["formula"]], data = dat, se.fit = FALSE, 
    conf.int = FALSE, times = 5))
o_auc <- map_dbl(o_auc_brier, ~ .x[["AUC"]][["score"]][["AUC"]])
o_brier <- map_dbl(o_auc_brier, ~ .x[["Brier"]][["score"]][["Brier"]][[2]])
# Calculate bootstrap-corrected AUC and Brier
boot_auc <- app_auc - mean(b_auc - o_auc)
boot_brier <- app_brier - mean(b_brier - o_brier)
\end{lstlisting}

\noindent
The bootstrap-corrected AUC is \verb|boot_auc=0.717| and the bootstrap-corrected Brier score is \\\verb|boot_brier=0.093|. Next, calculate the .632 performance. Begin by creating test datasets for each bootstrap sample, consisting of records from the original imputed dataset that were not included in the respective bootstrap sample. Then, evaluate the performance of the models fitted on each bootstrap sample using their corresponding test datasets. The .632 performance is calculated as 0.368 times the apparent performance plus 0.632 times the average performance across the test datasets:
\begin{lstlisting}[language=R]
# Obtain test datasets for each bootstrap sample
test_dat <- boot_imp_dat |> map(~ dat |> filter(!(id %in%
    unique(.x[["id"]]))))
# Calculate performance in each test dataset
test_auc_brier <- map2(test_dat, b_mod, ~ Score(list("b_fit" = .y),
    formula = app_mod[["formula"]], data = .x, se.fit = FALSE,
    conf.int = FALSE, times = 5))
test_auc <- map_dbl(test_auc_brier, ~ .x[["AUC"]][["score"]][["AUC"]])
test_brier <- map_dbl(test_auc_brier, ~ .x[["Brier"]][["score"]][["Brier"]][[2]])
# Calculate .632 AUC and Brier
boot632_auc <- .368 * app_auc + .632 * mean(test_auc)
boot632_brier <- .368 * app_brier + .632 * mean(test_brier)
\end{lstlisting}

\noindent
The .632 AUC is \verb|boot632_auc=0.717| and the .632 Brier score is \verb|boot632_brier=0.093|. Finally, calculate the .632+ performance. First, fix the ``no information performance'' at 0.5 for the AUC and 0.25 for the Brier score, representing the value of the performance metric for a non-informative predictive model. The relative overfitting rate is calculated as the difference between the average performance in the test datasets minus the apparent performance, divided by the difference between the ``no information performance'' value and the apparent performance. Next, define weights as 0.632 divided by 1 minus 0.368 times the relative overfitting rate. Finally, the .632+ performance is calculated as 1 minus the weight times the apparent performance plus the weight times the average performance in the test datasets:
\begin{lstlisting}[language=R]
# Fix "no information performance" for each metric
gamma_auc <- 0.5
gamma_brier <- 0.25
# Define relative overfitting rate
r_auc <- (mean(test_auc) - app_auc) / (gamma_auc - app_auc)
r_brier <- (mean(test_brier) - app_brier) / (gamma_brier - app_brier)
# Define weights
w_auc <- .632 / (1 - .368 * r_auc)
w_brier <- .632 / (1 - .368 * r_brier)
# Calculate the .632+ AUC and Brier
boot632plus_auc <- (1 - w_auc) * app_auc + w_auc * mean(test_auc)
boot632plus_brier <- (1 - w_brier) * app_brier + w_brier * mean(test_brier)
\end{lstlisting}

\noindent
The .632+ AUC is \verb|boot632plus_auc=0.717| and the .632+ Brier score is \verb|boot632plus_brier=0.093|.

From the guided example, we find that our original model was not very overfit, as the difference between the apparent performance metrics and the bootstrap-corrected performance metrics is somewhat small. We also had little overestimation of performance based on the inclusion of, on average, $63.2\%$ of patients from the original data in each bootstrap sample, as reflected in the similarly small difference between the bootstrap-corrected performance metrics and the .632 performance metrics. The .632+ performance metrics are identical to the .632 performance metrics, which is expected in this setting given the lack of strong overfit in the original model.

\section{Simulation study}\label{sec5}

The simulation study employed a multivariable model with 11 covariates across 54 scenarios, considering two sample sizes (750, 3500), nine missing data patterns, and three imputation methods as described in Section~\ref{sec3d}. The nine missing data patterns included: $x_1$ missing at (A) 5\%, (B) 15\%, (C) 60\%; $(x_1, x_3, x_4)$ missing at (D) (5\%, 5\%, 5\%), (E) (5\%, 15\%, 30\%), and (F) (15\%, 30\%, 60\%); and $(x_1, x_3, x_4, x_7, x_{10}, x_{11})$ missing at (G) (5\%, 5\%, 5\%, 5\%, 5\%, 5\%), (H) (5\%, 5\%, 15\%, 15\%, 30\%, 30\%), and (I) (15\%, 15\%, 30\%, 30\%, 60\%, 60\%). To reflect the complex joint missingness seen in real-world data, we applied pre-specified marginal and joint missingness proportions (Table~\ref{tab4}), following the specific missing data mechanisms for each covariate as outlined in Table~\ref{tab3}.

We compared the performance of complete case analysis (CC) with bootstrapping followed by deterministic imputation (BI). For each method, we generated 500 bootstrap samples. The bias of the various performance metrics defined in Section~\ref{sec2} was calculated as the difference between the performance metric on the full data and the performance metric employing CC and BI approaches. The bias of individual patient predicted probabilities was calculated as the difference in the average predicted probability in the full data and the average predicted probability employing CC and BI approaches. Both the predicted probabilities and the time-dependent AUCs are calculated at 5 years. 1000 simulated datasets were generated.

\begin{figure}[!ht]
\centering
\caption{Distribution of bias (y-axis) of AUC estimators (panels) comparing the CC approach and the BI approach with three imputation methods (color) when the sample size is 750 across missing data patterns (x-axis).\label{fig-750-auc}}
\includegraphics[scale = 0.9]{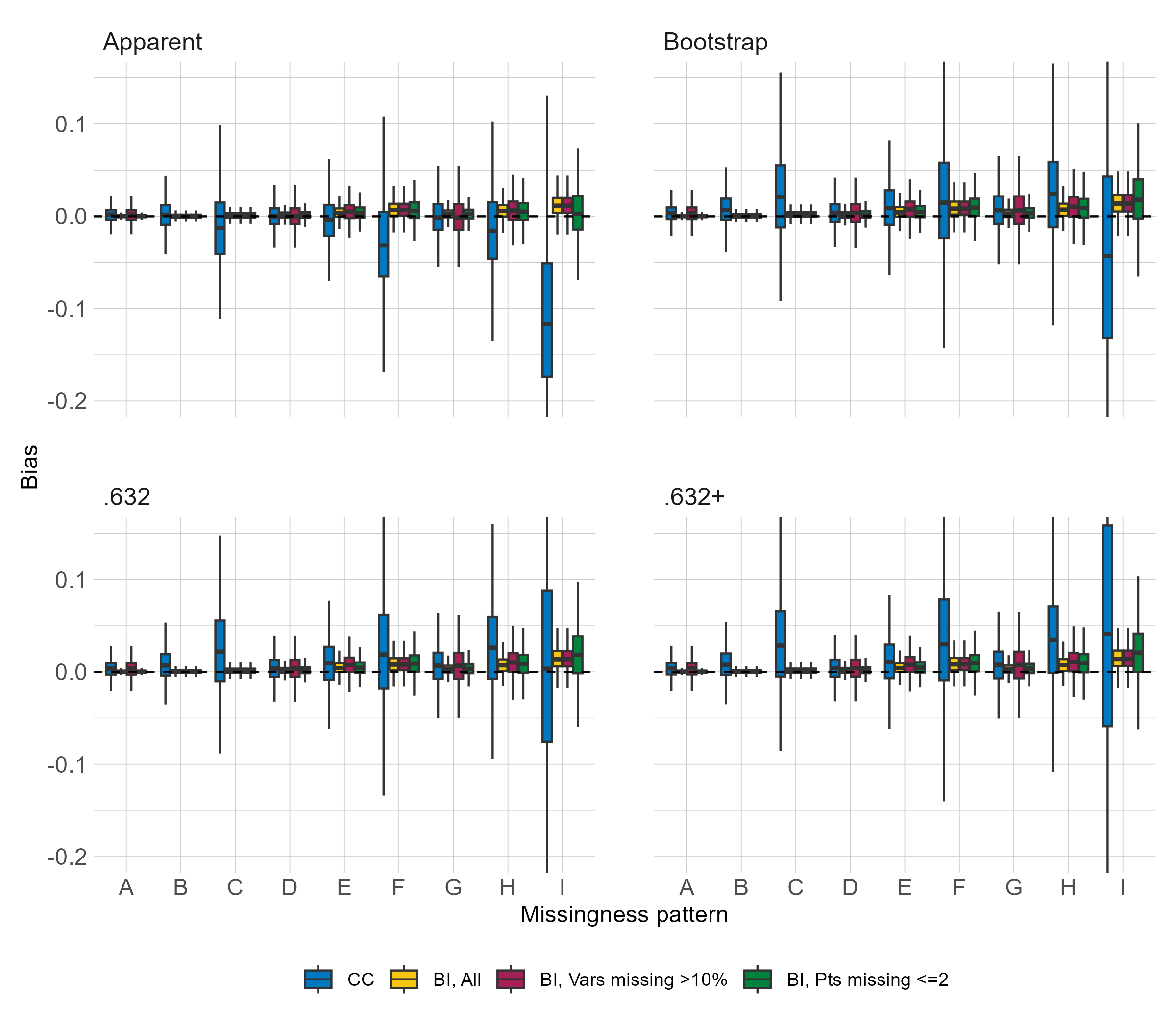}
\end{figure}

\begin{figure}
\centering
\caption{Distribution of bias (y-axis) of AUC estimators (panels) comparing the CC approach and the BI approach with three imputation methods (color) when the sample size is 3500 across missing data patterns (x-axis).\label{fig-3500-auc}}
\includegraphics[scale = 0.9]{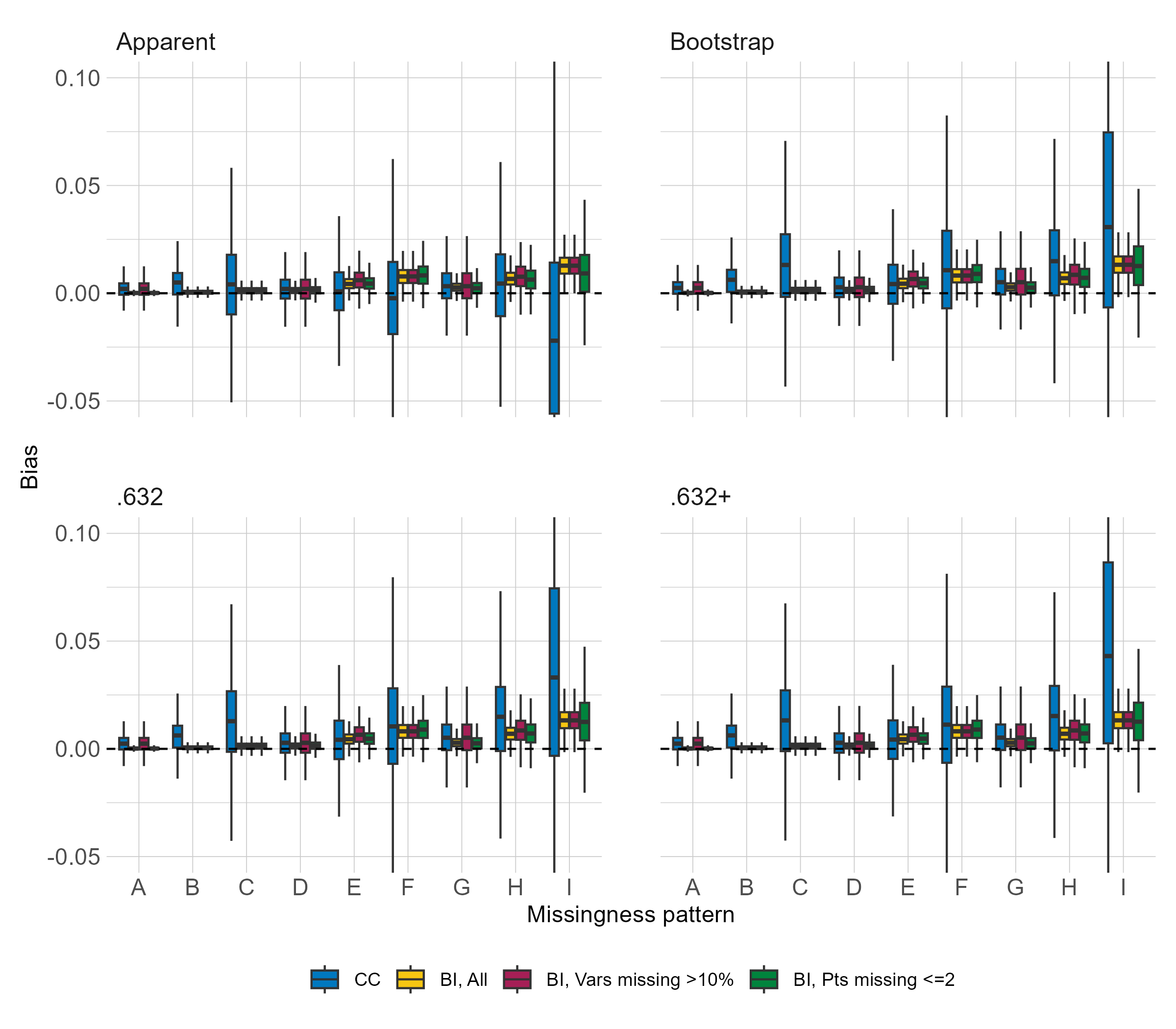}
\end{figure}

\begin{figure}
\centering
\caption{Distribution of bias (y-axis) of Brier score estimators (panels) comparing the CC approach and the BI approach with three imputation methods (color) when the sample size is 750 across missing data patterns (x-axis).\label{fig-750-brier}}
\includegraphics[scale = 0.9]{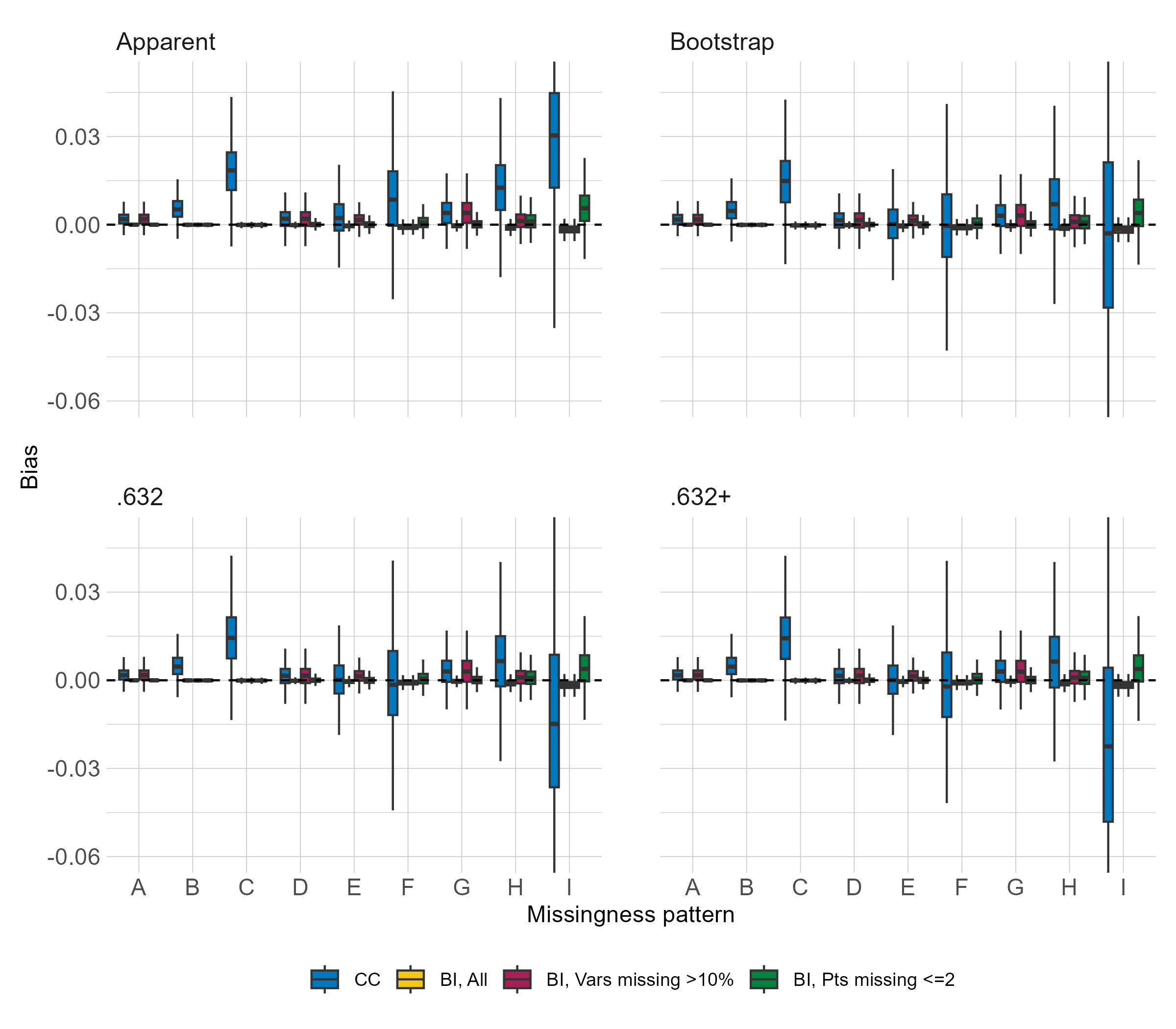}
\end{figure}

\begin{figure}
\centering
\caption{Distribution of bias (y-axis) of Brier score estimators (panels) comparing the CC approach and the BI approach with three imputation methods (color) when the sample size is 3500 across missing data patterns (x-axis).\label{fig-3500-brier}}
\includegraphics[scale = 0.9]{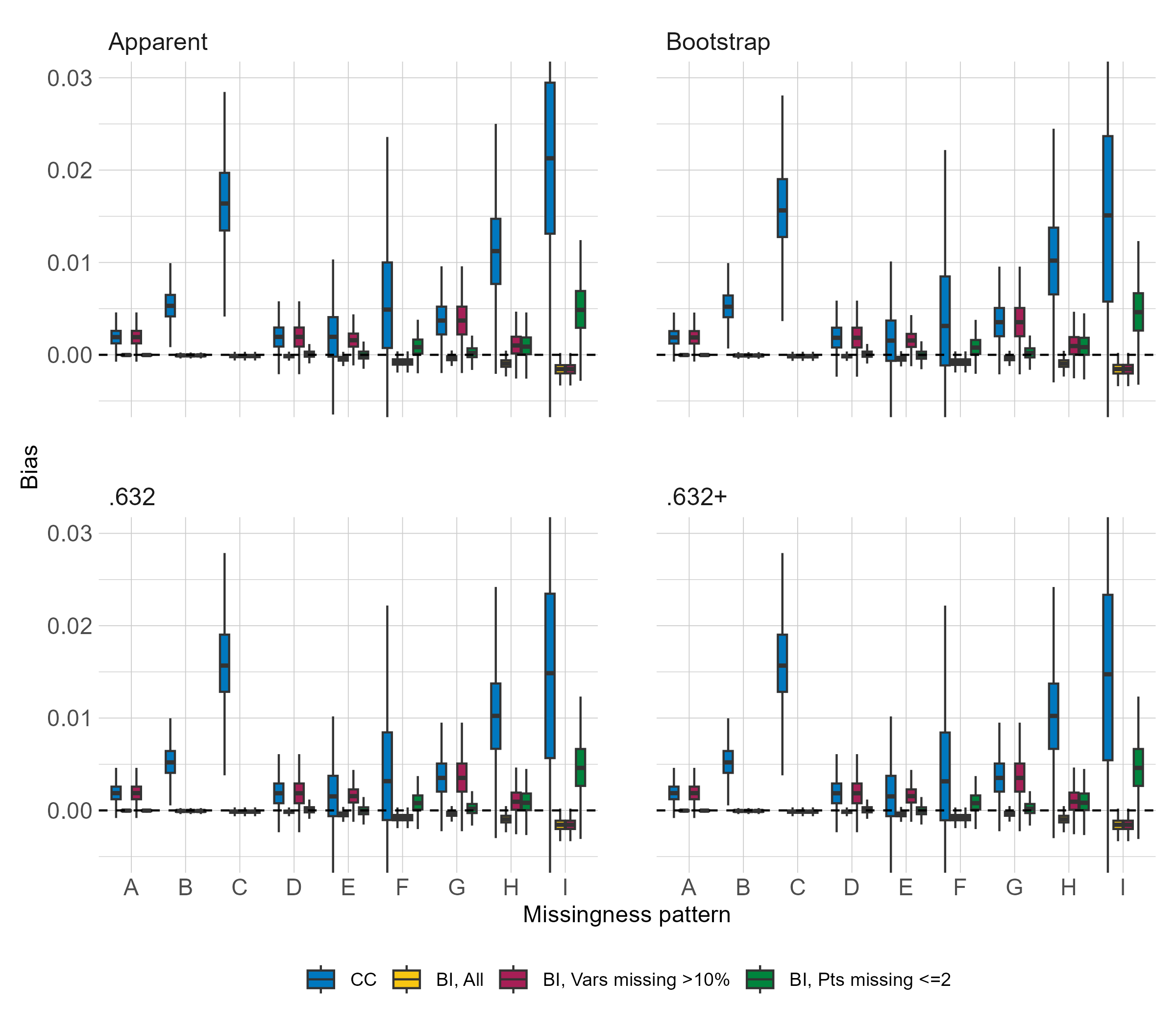}
\end{figure}

The distribution of bias across the nine missing data patterns for each of the four performance metrics are shown in Figures~\ref{fig-750-auc}, \ref{fig-3500-auc}, \ref{fig-750-brier}, and \ref{fig-3500-brier}. Each figure presents results for a combination of a single sample size for either the AUC or the Brier score. When the sample size is 750 and the approach is CC, the complete data were too sparse to successfully fit the multivariable model of interest in 0.3\%, 1.2\%, 0.4\%, and 57.7\% of simulated datasets for missing data patterns C, F, H, and I, respectively. These are missing data patterns with either 60\% missingness for a given covariate or with missingness in 6 covariates, some of which are missing in $>5\%$ of participants. Even when the sample size is 3500, the CC approach is unable to fit the multivariable model of interest for missing pattern I in 0.3\% of simulated datasets. The figures are based on results from simulated datasets where the multivariable model of interest was successfully fitted, resulting in smaller sample sizes for certain settings.

The CC approach always has more bias and more variability in the bias, as compared to the BI approach with any of the imputation methods. For the BI approach, the bias is equal or lower across almost all settings when we impute all versus only imputing covariates missing $>10\%$ or only imputing for participants with $\leq2$ missing covariates, the approach with the highest bias. This holds true even for missing data patterns with higher levels of missingness within a covariate or higher numbers of missing data. With respect to the missing data pattern, the bias is larger for the more extreme missing data patterns. With the CC approach, regardless of metric or sample size, missingness patterns C, F, H, and I have the largest bias and also the most variability of bias (though recall the variability is based on a smaller sample size due to model fitting failures). With the BI approach, regardless of metric or sample size, the bias is highest in settings with multiple missing covariates, and increases as the proportion of missingness increases, so that the highest bias occurs in settings E, F, H, and I.

\begin{figure}
\centering
\caption{Distribution of average individual prediction bias (y-axis) by sample size (columns) for the CC approach and the BI approach with three imputation methods (color) across missing data patterns (x-axis).\label{fig-pred-bias}}
\includegraphics[scale = 0.9]{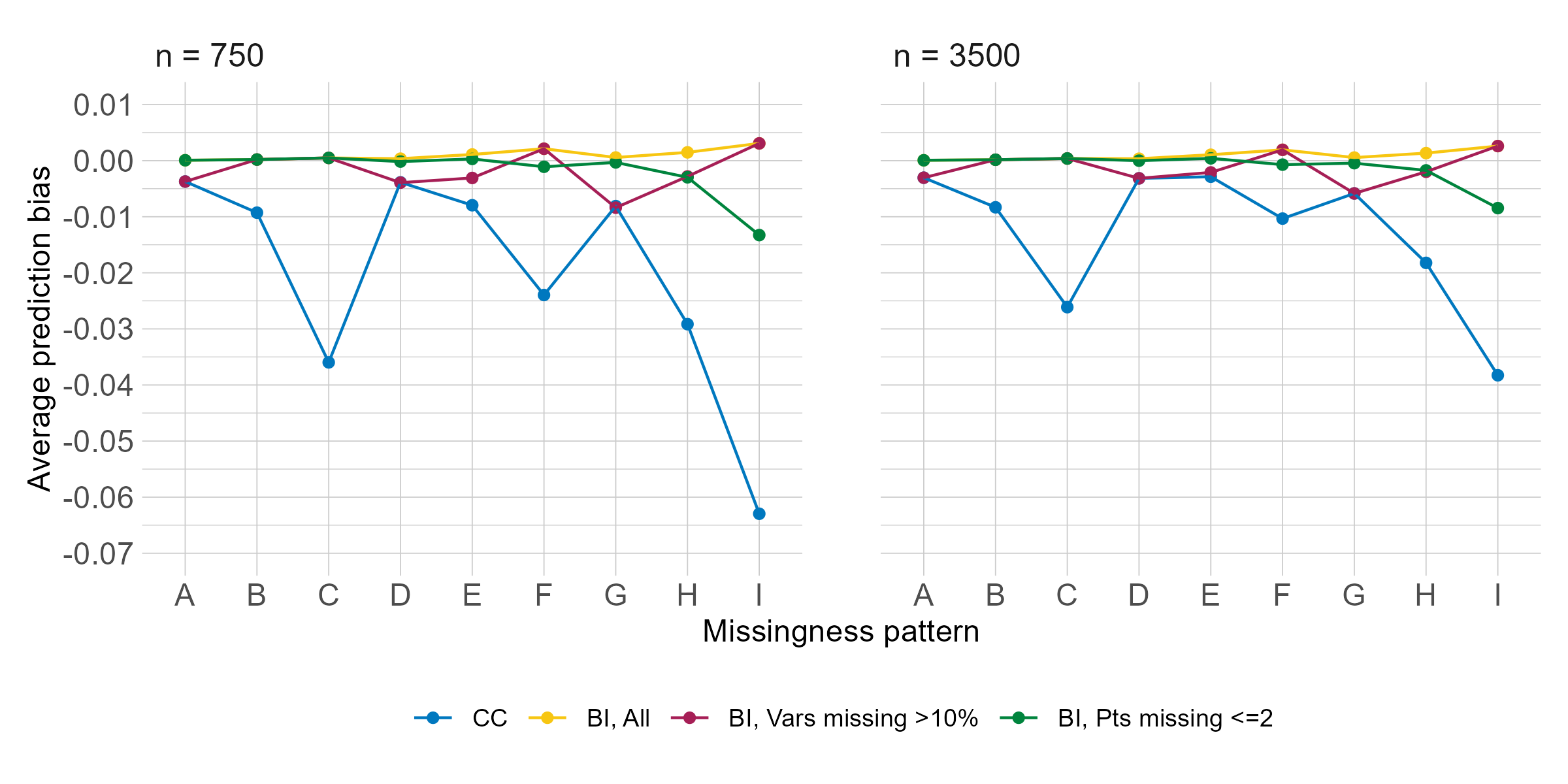}
\end{figure}

The distribution of average individual prediction bias is shown in Figure~\ref{fig-pred-bias}. We find that the average individual prediction bias is quite low across all settings, and tends to be in a negative direction, indicating that the predicted risks using these missing data approaches are smaller than the predicted risks in the true full data. In addition, the average individual prediction bias using the BI approach is always equal to or less than the average individual prediction bias using the CC approach. When using the BI approach, the bias is equal or lower across almost all scenarios when we impute all versus only imputing covariates missing $>10\%$ or only imputing for participants with $\leq2$ missing covariates. For missing data patterns A, D, E, and G, when the approach is BI, imputing only covariates missing $>10\%$ has the highest bias whereas in missing data pattern I, imputing for participants with $\leq2$ missing covariates has the highest bias.

The R functions and code to run all simulation studies included in this section are available on GitHub at \url{https://github.com/zabore/manuscript-code-repository/tree/master/Mi-Zabor_bootstrap-impute-predict-tutorial}.

\section{Discussion}

The development of clinical risk prediction models has proliferated in the past decade, as data sources become more available and there is increasing interest in personalized medicine. In this paper, we have provided practical guidance on the most appropriate approach to handling missing data when the goal is to build a clinical risk prediction model with internal validation in the presence of missing data. Deterministic regression imputation enables predictions for future patients by excluding the outcome from the imputation model. Missing covariates for future patients can be easily imputed using regression coefficients derived from the imputation model. Therefore, the results of a clinical prediction model developed and validated in this way could easily be published for future use or deployed in an online risk prediction calculator. The step-by-step tutorial, coupled with a detailed discussion of the rationale behind the proposed approach, aims to guide future statisticians and clinicians in developing clinical risk prediction models.

The results of our simulation studies reveal that bootstrapping followed by deterministic imputation, where all missing values are imputed, results in the least biased estimates of prediction model performance as measured by either the AUC or the Brier score, and in the least biased estimates of individual risk predictions. This paper did not aim to directly compare performance metrics, as that work has been done previously \cite{Efron1997, Iba2021}. Rather, the aim was to compare approaches to handling missing data in the context of commonly used performance metrics.

Even in settings with low levels of missing data, as low as 5\% in a single covariate, we found that bootstrapping followed by imputation resulted in lower bias than complete case analysis. In settings with high levels of missing data, such as one covariate missing for 60\% of participants, three covariates missing for 15\%, 30\% and 60\% of participants, and six covariates missing in 5\%, 5\%, 15\%, 15\%, 30\%, 30\% or in 15\%, 15\%, 30\%, 30\%, 60\%, 60\%, complete case analysis was not always feasible, as the model might fail to converge, or if it did converge, it could be severely overfit. The feasibility and reliability of complete case analysis depend on the total sample size, the complete case sample size, and the number of covariates included in the final prediction model. In such scenarios, imputing missing values is necessary to conduct the desired analysis. Our findings indicate that even in settings with high levels of missingness, and with a total sample size as small as 750, using bootstrapping followed by imputing all missing values resulted in relatively low bias. 

There are many reasons why covariate data may be missing from datasets used to develop and internally validate a clinical risk prediction model. A common cause is the use of retrospective datasets, where variables of interest were not routinely collected but were instead captured incidentally during routine clinical care. In such cases, the missing data are likely to meet the assumption of being missing at random (MAR). At other times, a single covariate may have high levels of missingness for specific reasons, such as resulting from an expensive, difficult-to-access, or invasive test. In such cases, the assumption of MAR may not be appropriate. Careful consideration is needed to determine whether other measured covariates could provide insight into the factors influencing whether the test was performed. In the context of external validation, a variable routinely collected at the institution providing the development data might not have been routinely collected at the institution providing the validation data. Reasons for missing data are complex and varied, and here we do not even address the critical issue of individuals who are entirely absent from datasets due to lack of access to routine healthcare.

The development and validation of clinical risk prediction models require careful thought on many levels \cite{Zabor2018}. In this work, we have focused on one critical aspect: handling missing data in the context of internal validation. It is crucial that everyone involved in developing and validating clinical risk prediction models put careful thought into the choices being made that may impact results. Above all, the primary objective should always be to develop a model that is practical and reliable for use by both patients and clinicians.

\addcontentsline{toc}{section}{References}

\bibliographystyle{unsrt}
\bibliography{MI-bootstrap-prediction-refs}

\clearpage

\section*{Tables and Figures}

\begin{table}[ht]
\caption{True hazard ratios.\label{tab1}}
\begin{tabular}{ l l l }
\hline
\textbf{Covariate} & \textbf{Type} & \textbf{Hazard ratio} \\
\hline
$X_1$ & Binary & 0.80 \\
$X_2$ & Continuous & 1.05 \\
$X_3$ & Continuous & 1.25 \\
$X_4$ & Binary & 1.54 \\
$X_5$ & Binary & 1.18 \\
$X_6$ & Continuous & 1.45 \\
$X_7$ & Binary & 1.10 \\
$X_8$ & Binary & 0.76 \\
$X_9$ & Binary & 0.64 \\
$X_{10}$ & Binary & 1.25 \\
$X_{11}$ & Binary & 0.48 \\
\hline
\end{tabular}
\end{table}

\begin{table}[ht]
\small
\caption{Covariance matrix for simulated data.\label{tab2}}
\begin{tabular}{lrrrrrrrrrrr}
\hline
 & $X_1$ & $X_2$ & $X_3$ & $X_4$ & $X_5$ & $X_6$ & $X_7$ & $X_8$ & $X_9$ & $X_{10}$ & $X_{11}$ \\ 
\hline
$X_1$ & 0.2370 &  &  &  &  &  &  &  &  &  &  \\ 
  $X_2$ & -1.3349 & 196.2990 &  &  &  &  &  &  &  &  &  \\ 
  $X_3$ & 0.0812 & 0.6471 & 1.0967 &  &  &  &  &  &  &  &  \\ 
  $X_4$ & 0.0247 & -0.6314 & 0.0680 & 0.2441 &  &  &  &  &  &  &  \\ 
  $X_5$ & 0.0298 & -0.0759 & 0.0452 & 0.0063 & 0.1302 &  &  &  &  &  &  \\ 
  $X_6$ & 0.0134 & 0.3167 & 0.0063 & -0.0089 & 0.0070 & 0.0558 &  &  &  &  &  \\ 
  $X_7$ & 0.0280 & -0.7944 & 0.0740 & 0.0468 & 0.0214 & -0.0015 & 0.2427 &  &  &  &  \\ 
  $X_8$ & -0.0069 & 0.0156 & -0.0190 & -0.0581 & -0.0026 & 0.0050 & -0.0119 & 0.1146 &  &  &  \\ 
  $X_9$ & 0.0039 & -0.1261 & -0.0080 & 0.0484 & 0.0016 & -0.0026 & 0.0131 & -0.0318 & 0.1408 &  &  \\ 
  $X_{10}$ & 0.0002 & 0.0294 & -0.0147 & 0.0003 & -0.0001 & 0.0003 & 0.0017 & 0.0001 & 0.0050 & 0.1369 &  \\ 
  $X_{11}$ & 0.0223 & -0.8057 & -0.0075 & -0.0139 & 0.0043 & -0.0000 & 0.0012 & 0.0086 & -0.0267 & 0.0014 & 0.0986 \\ 
\hline
\end{tabular}
\end{table}

\begin{table}[!ht]
\caption{Fixed probability of missingness parameters.\label{tab3}}
\begin{tabular}{ l l l l}
\hline
\textbf{Missing covariate} & \textbf{$k_j$} & \textbf{$l_j$}  & \textbf{$\gamma_{2j}$}\\
\hline
$X_1$ & --- & $2$ & $\log(1.05)$ \\
$X_3$ & $1$ & $5$ & $\log(0.80)$ \\
$X_4$ & $3$ & $6$ & $\log(0.70)$ \\
$X_7$ & $4$ & $8$ & $\log(0.90)$ \\
$X_{10}$ & $5$ & $9$ & $\log(0.60)$ \\
$X_{11}$ & $10$ & $2$ & $\log(1.05)$ \\
\hline
\end{tabular}
\end{table}

\begin{table}[!ht]
\caption{Marginal and joint missing data proportions.\label{tab4}}
\begin{tabular}{rrr}
\hline
Marginal missingness $X_i$ & Marginal missingness $X_j$ & Joint missingness $(X_i, X_j)$ \\ 
\hline
0.05 & 0.05 & 0.01 \\ 
  0.05 & 0.15 & 0.02 \\ 
  0.15 & 0.15 & 0.05 \\ 
  0.15 & 0.30 & 0.07 \\ 
  0.30 & 0.30 & 0.10 \\ 
  0.30 & 0.60 & 0.20 \\ 
  0.60 & 0.60 & 0.40 \\ 
\hline
\end{tabular}
\end{table}

\vspace{15em}

\end{document}